\documentclass[aps,prd,onecolumn,groupedaddress,showpacs,nofootinbib,amssymb]{revtex4}
\usepackage[dvips]{graphicx}
\usepackage{amssymb}
\usepackage{amsmath}
\usepackage{graphicx,,color}
\usepackage{amsfonts}
\usepackage{bm}
\usepackage{cancel}
\usepackage{comment}

\newcommand\be{\begin{equation}}
\newcommand\ee{\end{equation}}

\allowdisplaybreaks[4]

\begin{document}

\tolerance=5000

\title{Maximum Baryon Masses for Static Neutron Stars in $f(R)$ Gravity}
\author{A.V. Astashenok$^{1}$, S. Capozziello$^{2,3,4}$\thanks{capozzie@na.infn.it}
S.D.~Odintsov,$^{5,6}$\,\thanks{odintsov@ieec.uab.es}
V.K.~Oikonomou,$^{7,8}$\,\thanks{v.k.oikonomou1979@gmail.com}}
\affiliation{$^{1)}$ Institute of Physics, Mathematics and IT, I.
Kant Baltic Federal University, Kaliningrad, 236041 Russia,\\
$^{2)}$ Dipartimento di Fisica, ''E. Pancini'' Universit`a
''Federico II'' di Napoli, Compl. Univ. Monte S. Angelo
Ed. G, Via Cinthia, I-80126 Napoli, Italy,\\
$^{3)}$INFN Sez. di Napoli, Compl. Univ. Monte S. Angelo Ed. G,
Via Cinthia, I-80126 Napoli, Italy,
\\
$^{4)}$ Scuola Superiore Meridionale, Largo S. Marcellino 10, I-80138 Napoli, Italy,\\
$^{5)}$ ICREA, Passeig Luis Companys, 23, 08010 Barcelona, Spain\\
$^{6)}$ Institute of Space Sciences (IEEC-CSIC) C. Can Magrans
s/n,
08193 Barcelona, Spain\\
 $^{7)}$ Department of Physics,
Aristotle University of Thessaloniki, Thessaloniki 54124,
Greece\\
$^{8)}$ Laboratory for Theoretical Cosmology, Tomsk State
University of Control Systems and Radioelectronics (TUSUR), 634050
Tomsk,
Russia,\\
}

\date{\today}
\tolerance=5000

\begin{abstract}
We investigate the upper mass limit predictions of the baryonic
mass for static neutron stars in the context of $f(R)$ gravity. We
use the most popular $f(R)$ gravity model, namely the $R^2$
gravity, and calculate the maximum baryon mass of static neutron
stars adopting several realistic equations of state and one ideal
equation of state, namely that of  causal limit. Our motivation is
based on the fact that neutron stars with baryon masses larger
than the maximum mass for static neutron star configurations
inevitably collapse to black holes. Thus with our analysis,  we
want further to enlighten the predictions for the maximum baryon
masses of static neutron stars in $R^2$ gravity, which, in turn,
further strengthens our understanding of the mysterious mass-gap
region. As we show, the baryon masses of most of the equations of
states studied in this paper, lie in the lower limits of the
mass-gap region $M\sim 2.5-5 M_{\odot}$, but intriguingly enough,
the highest value of the maximum baryon masses we found is of the
order of $M\sim 3 M_{\odot}$. This upper mass limit also appears
as a maximum static neutron star gravitational mass limit in other
contexts. Combining the two results which refer to baryon and
gravitational masses, we point out that the gravitational mass of
static neutron stars cannot be larger than three solar masses,
while based on maximum baryon masses results of the present work,
we can conspicuously state that it is highly likely the lower mass
limits of astrophysical black holes  in the range of $M\sim 2.5-3
M_{\odot}$. This, in turn, implies that maximum neutron star
masses in the context of $R^2$ gravity are likely to be in the
lower limits of the range of $M\sim 2.4-3 M_{\odot}$. Hence our
work further supports the General Relativity claim that neutron
stars cannot have gravitational masses larger than $3$$M_{\odot}$
and then, to explain observations comparable or over this limit,
we need alternative extensions of General Relativity, other than
$f(R)$ gravity.
\end{abstract}


\maketitle

\section{Introduction}

Currently and for the next 10-15 years, gravitational waves and
neutron stars are in the focus of the scientific community. As it
seems, the Large Hadron Collider at CERN indicates than new
physics may lie well above 15 TeV center of mass, thus
contemporary science is focused on neutron star (NS) physics (for
an important stream of reviews and textbooks see for example
\cite{Haensel:2007yy,Friedman:2013xza,Baym:2017whm,Lattimer:2004pg,Olmo:2019flu})
and astrophysical objects merging, which may provide insights to
fundamental physics problems. Indeed, neutron stars (NSs) have
multiple correlations with various physics research areas, like
nuclear physics
\cite{Lattimer:2012nd,Steiner:2011ft,Horowitz:2005zb,Watanabe:2000rj,Shen:1998gq,Xu:2009vi,Hebeler:2013nza,Mendoza-Temis:2014mja,Ho:2014pta,Kanakis-Pegios:2020kzp},
high energy physics
\cite{Buschmann:2019pfp,Safdi:2018oeu,Hook:2018iia,Edwards:2020afl,Nurmi:2021xds},
modified gravity
\cite{Astashenok:2020qds,Astashenok:2021xpm,Astashenok:2021peo,Capozziello:2015yza,Astashenok:2014nua,Astashenok:2014pua,Astashenok:2013vza,Arapoglu:2010rz,Panotopoulos:2021sbf,Lobato:2020fxt,Oikonomou:2021iid,Odintsov:2021nqa,Odintsov:2021qbq}
and relativistic astrophysics
\cite{Bauswein:2020kor,Vretinaris:2019spn,Bauswein:2020aag,Bauswein:2017vtn,Most:2018hfd,Rezzolla:2017aly,Nathanail:2021tay,Koppel:2019pys}.
Apart from the physical implications of isolated neutron stars,
surprises for fundamental physics may arise from the merging of
astrophysical objects, the mysteries of which are analyzed by the
LIGO-Virgo collaboration. Already the GW170817 event
\cite{TheLIGOScientific:2017qsa} has changed the way of thinking
in theoretical cosmology indicating that the gravitational waves
propagate with a speed equal to that of light. The physics of
astrophysical gravitational waves, and more importantly that of
primordial gravitational waves is expected to change the way of
thinking, or to verify many theoretical proposals in theoretical
cosmology. Future collaborations like the Einstein Telescope Hz
KHz frequencies \cite{Hild:2010id}, the LISA Space-borne Laser
Interferometer Space Antenna \cite{Baker:2019nia,Smith:2019wny}
the BBO \cite{Crowder:2005nr,Smith:2016jqs}, DECIGO
\cite{Seto:2001qf,Kawamura:2020pcg} and finally the SKA (Square
Kilometer Array) Pulsar Timing Arrays at frequencies $10^{-8}$Hz
\cite{Bull:2018lat} are expected to shed new light to fundamental
high energy physics problems, with most of these telescopes
revealing the physics of the radiation domination era. As we
already stated, in the next 10-15 years, particle physics,
theoretical cosmology and theoretical astrophysics will heavily
rely to gravitational wave and NSs observations. Although it seems
that things are more or less settled with theoretical
astrophysics, a recent observation \cite{Abbott:2020khf} has cast
doubt on the maximum mass issue of NSs, and, in parallel, it
indicated that alternative astrophysical objects, like strange
stars, may come into play in the near future. Although it is quite
early phenomenologically speaking, for exotic stars to be
discovered, it is a realistic possibility. Then, if exotic objects
are not yet fully phenomenologically supported, the problem with
the observation \cite{Abbott:2020khf} is that it is probable to
find NSs with masses in the mass-gap region $M\sim 2.5-5
M_{\odot}$. This possibility is sensational and it raises the
fundamental question inherent to the maximum mass problem of NSs,
which is, what is the lowest mass of astrophysical black holes. In
the context of General Relativity (GR), non-rotating neutron stars
with masses in the mass-gap region can only be described by
ultra-stiff equations of state (EoS), thus it is rather hard to
describe them without being in conflict with the GW170817 results.
Modified and extended gravity in its various forms
\cite{reviews1,reviews2,reviews3,reviews4,book,reviews5,reviews6,dimo}
can provide a clear cut description for large mass NSs
\cite{Pani:2014jra,Doneva:2013qva,Horbatsch:2015bua,Silva:2014fca,Chew:2019lsa,Blazquez-Salcedo:2020ibb,Motahar:2017blm,Oikonomou:2021iid,Odintsov:2021nqa,Odintsov:2021qbq}
see also Refs.
\cite{Astashenok:2020qds,Astashenok:2021peo,Astashenok:2021xpm}
for recent descriptions of the GW190814 event, and thus serves as
a cutting edge probable description of nature in limits where GR
needs to be supplemented by a Occam's razor compatible theory.

Motivated by the fundamental and inherently related questions,
which is the maximum mass of NSs and what is the lowest mass of
astrophysical black holes, in this work we shall approach these
issues by calculating the maximum baryonic mass of NSs in the
context of $f(R)$ gravity. We shall focus our analysis on one of
the most popular to date  models of $f(R)$ gravity, the $R^2$
model, and we shall use a large variety of EoSs for completeness.
Our approach will yield an indirect way to answer the question
what is the lowest limit of astrophysical black holes, since it is
known that NSs with baryonic masses larger than the maximum static
limit will eventually collapse into black holes. Thus by knowing
the maximum baryonic mass, one may easily calculate the universal
gravitational mass-baryonic mass relation (see Ref.
\cite{Gao:2019vby} for the GR case) for the corresponding $R^2$
gravity theory, and can eventually find the gravitational mass of
the NSs corresponding to the maximum baryon mass. Having the
maximum gravitational mass available, one answers both questions
discussed in this paragraph, since NSs with baryonic masses larger
than the maximum allowed, will collapse into black holes. Thus our
work paves the way towards answering the two fundamental questions
related with the mysterious mass gap regions. The focus in this
work is to find the maximum baryon masses for $R^2$ gravity NSs
for various EoSs, and this is the first toward revealing  the
mass-gap region.

\section{Maximum Baryon Masses in $f(R)$ Gravity}

Let us calculate numerically the maximum
baryonic mass of static NSs in the context of $f(R)$ gravity and
specifically for the $R^2$ model in the Jordan frame. We shall use
several different phenomenological EoSs and our main aim is to
pave the way towards answering the question what is the maximum
gravitational mass that neutron stars can have. At the same time,
if this question is answered, one may also have a hint on the
question which is the lowest mass that astrophysical black holes
can have. Before getting to the details of our analysis, we
provide here an overview of the treatment of spherically symmetric
compact objects in the context of Jordan frame $f(R)$ gravity, and
the Tolman-Oppenheimer-Volkoff (TOV) equations.

The $f(R)$ gravity action in the Jordan frame is the following,
\begin{equation}\label{action}
{\cal A}=\frac{c^4}{16\pi G}\int d^4x \sqrt{-g}\left[f(R) + {\cal
L}_{{\rm matter}}\right]\,,
\end{equation}
with $g$ denoting the metric tensor determinant and ${\cal L}_{\rm
matter}$ denotes the  Lagrangian of the perfect matter fluids that
are present. Upon variation of the action (\ref{action}) with
respect to the metric tensor $g_{\mu\nu}$, the field equations are
obtained \cite{reviews4},
\begin{equation}
\frac{df(R)}{d R}R_{\mu\nu}-\frac{1}{2}f(R)
g_{\mu\nu}-\left[\nabla_{\mu} \nabla_{\nu} - g_{\mu\nu}
\Box\right]\frac{df(R)}{dR}=\frac{8\pi G}{c^{4}} T_{\mu \nu },
\label{field_eq}
\end{equation}
for a general metric $g_{\mu \nu}$, where
$\displaystyle{T_{\mu\nu}=
\frac{-2}{\sqrt{-g}}\frac{\delta\left(\sqrt{-g}{\cal
L}_m\right)}{\delta g^{\mu\nu}}}$ stands for the energy-momentum
tensor of the perfect matter fluids present. We shall consider
static NSs, which are described by the following
spherically symmetric metric,
\begin{equation}
    ds^2= e^{2\psi}c^2 dt^2 -e^{2\lambda}dr^2 -r^2 (d\theta^2 +\sin^2\theta d\phi^2),
    \label{metric}
\end{equation}
where $\psi$ and $\lambda$ are arbitrary functions with radial
dependence only. The energy momentum tensor for the perfect matter
fluid describing the NS is,
$T_{\mu\nu}=\mbox{diag}(e^{2\psi}\rho c^{2}, e^{2\lambda}p, r^2 p,
r^{2}p\sin^{2}\theta)$ where $\rho$ denotes the energy-matter density and
$p$ stands for the pressure \cite{weinberg}. By using the
contracted Bianchi identities, one can obtain the equations for
the stellar object, by also implementing the hydrostatic
equilibrium condition,
\begin{equation} \nabla^{\mu}T_{\mu\nu}=0\,,
\label{bianchi}
\end{equation}
which, in turn,  yields  the Euler conservation equation,
\begin{equation}\label{hydro}
    \frac{dp}{dr}=-(\rho
    +p)\frac{d\psi}{dr}\,.
\end{equation}
Upon combining the metric \eqref{metric} and the field  Eqs.
(\ref{field_eq}), we obtain the equations governing the behavior
of the functions $\lambda$ and $\psi$ inside and outside the
compact object, which are \cite{capquark},
\begin{eqnarray}
\label{dlambda_dr} \frac{d\lambda}{dr}&=&\frac{ e^{2 \lambda
}[r^2(16 \pi \rho + f(R))-f'(R)(r^2 R+2)]+2R_{r}^2 f'''(R)r^2+2r
f''(R)[r R_{r,r} +2R_{r}]+2 f'(R)}{2r \left[2 f'(R)+r R_{r}
f''(R)\right]},
\end{eqnarray}
\begin{eqnarray}\label{psi1}
\frac{d\psi}{dr}&=&\frac{ e^{2 \lambda }[r^2(16 \pi p -f(R))+
f'(R)(r^2 R+2)]-2(2rf''(R)R_{r}+ f'(R))}{2r \left[2 f'(R)+r R_{r}
f''(R)\right]}\, ,
\end{eqnarray}
with the prime in Eqs. (\ref{dlambda_dr}) and (\ref{psi1})
denoting differentiation with respect to the function $R(r)$, that
is $f'(R)=\frac{d f}{d R}$. The above set of differential
equations constitute the $f(R)$ gravity TOV equations, and by
using $f(R)=R$ one obtains the standard  TOV equations of GR
\cite{rezzollazan,landaufluid}. In addition to the above TOV
equations, for $f(R)$ gravity, the TOV equations are also
supplemented by the following differential equation,
\begin{equation}\label{TOVR}
\frac{d^2R}{dr^2}=R_{r}\left(\lambda_{r}+\frac{1}{r}\right)+\frac{f'(R)}{f''(R)}\left[\frac{1}{r}\left(3\psi_{r}-\lambda_{r}+\frac{2}{r}\right)-
e^{2 \lambda }\left(\frac{R}{2} + \frac{2}{r^2}\right)\right]-
\frac{R_{r}^2f'''(R)}{f''(R)},
\end{equation}
which is obtained from the trace of Eqs.\eqref{field_eq} by
replacing the metric \eqref{metric}. The differential Eq.
(\ref{TOVR}) basically expresses the fact that the Ricci scalar
dynamically evolves in the context of $f(R)$ gravity, as the
radial coordinate $r$ changes.

Having presented the TOV equations, the focus  is now on
solving numerically them, namely Eqs. (\ref{hydro}),
(\ref{dlambda_dr}) and (\ref{psi1}) together with (\ref{TOVR}),
for the $R^2$ model,
\begin{equation}\label{frdef}
f(R)=R+\alpha R^2\, ,
\end{equation}
where the parameter $\alpha$ is expressed in units of
$r_g^2=4G^2M^2_\odot/c^4$ and $r_g$ is the Sun gravitational radius.
With regard to the EoS, we shall consider five phenomenological
and one ideal limiting case EoSs, specifically: a) the APR4 which
is a $\beta$-equilibrium EoS proposed by Akmal, Pandharipande and
Ravenhall (\cite{APR4}). b) The BHF which is a microscopic EoS of
dense $\beta$-stable nuclear matter obtained using realistic
two-body and three-body nuclear interactions denoted as
$N3LO\Delta$ + $N2LO\Delta_1$ \cite{BHF} derived in the framework
of chiral perturbation theory and including the $\Delta$(1232)
isobar intermediate state. This EoS has been derived using the
Brueckner-Bethe-Goldstone quantum many-body theory in the
Brueckner-Hartree-Fock approximation. c) The GM1 EoS, which is the
classical relativistic mean field parametrization GM1 \cite{GM}
for cold neutron star matter in $\beta$-equilibrium containing
nucleons and electrons. d) The QHC18, which is a phenomenological
unified EoS proposed in \cite{QHC18} and describes the crust,
nuclear liquid, hadron-quark crossover, and quark matter domains.
e) The SLy \cite{Douchin:2001sv}, which is a well known and
phenomenologically successful EoS. f) Finally, the limiting case
ideal EoS, called the causal limit EoS, in which case,
$$
P(\rho) = P_{u}(\rho_u) + (\rho-\rho_u)v_s^{2}.
$$
with $P_{u}$ and $\rho_u$ correspond to the pressure and density
of the well known segment of a low-density EoS at $\rho_u\approx
\rho_0$ where $\rho_0$ denotes the saturation density. We shall
assume that the low-density EoS is the SLy EoS and consider the case
when $v_s^{2}=c^2/3$. It is conceivable that the causal EoS is an
ideal limit, thus the resulting baryonic mass for static NSs that
will be obtained for this EoS, will serve as a true upper bound
for the baryonic masses NSs in $R^2$.

Now let us get into the core of our analysis. The clue point is
that NSs with baryon masses larger than $M_B^{max}$, that is,
\begin{equation}\label{mbax}
M_B>M_B^{max}\, ,
\end{equation}
will inevitably  collapse to black hole. Thus, in principle, by
knowing the the maximum baryonic mass for a specific EoS and a
specific theory can yield a first hint on where to find black
holes and what is, for sure, the upper limit of NSs, indirectly
though. Let us explain in detail these two syllogisms in some
detail, considering firstly the black hole syllogism although the
two are inherently related. If one knows the maximum baryonic mass
for a specific EoS and theory,  it can provide a hint on the
lowest mass of astrophysical black holes. Basically, we can
indirectly know where to start seeking for the lower mass limit of
astrophysical black holes, since $M_B^{max}$ is an ideal upper
limit that the gravitational masses of NSs cannot reach for sure.
Thus the lower limit of astrophysical black holes could be
$M_B^{max}$ because it is not possible to find NSs with such large
gravitational masses. On the other hand, and in the same line of
research, the gravitational masses of NSs can never be as large as
the maximum baryonic masses, thus it is expected to find them to
quite lower limiting values. Thus, in the NSs case, we know where
not to find NSs and seek them in quite lower values. In both
cases, the analysis would be perfectly supplemented by knowing for
a large number of EoSs and a large number of theories, the
theoretical universal relation between the baryonic and
gravitational NSs masses, as in \cite{Gao:2019vby}. However this
task is quite complicated and it will be addressed in more detail
in a future focused work. In this work, we aim to find hints on
where to start finding the lowest limit of astrophysical black
holes, and also to discover where not to find NSs, thus aiming in
providing another theoretical upper bound on static NSs masses.
This work could be considered as a theoretical complement of our
work on causal EoSs developed in Ref. \cite{Astashenok:2021peo}.
Remarkably, the two results seem to provide a quite interesting
result and may lead to an interesting conjecture.

Let us proceed by briefly recalling how to calculate the baryonic
mass for a static NS. The central values of the pressure and of
the mass of the NS are,
\begin{equation}\label{pressuremassatthecenter}
P(0) = P_c,\,\,\,m(0) = 0\, ,
\end{equation}
and near the center, the pressure and the mass of the NS behave
as,
\begin{equation}\label{nearcenterpressure}
P(r) \simeq P_c -(2\pi)(\epsilon_c+P_c) \left(
P_c+\frac{1}{3}\epsilon_c \right) r^2 + O(r^4)\, ,
\end{equation}
\begin{equation}\label{nearthecentermass}
m(r) \simeq \frac{4}{3}\pi\epsilon_cr^3 + O(r^4)\, .
\end{equation}
Considering the spherically symmetric spacetime (\ref{metric}),
the gravitational mass of the NS is,
\begin{equation}\label{gravitationalmass}
M= \int_0^R 4\pi r^2\epsilon dr\, ,
\end{equation}
or equivalently,
\begin{equation}\label{gravimassalternative}
 M= \int_0^R 4\pi r^2 e^{(\psi+\lambda)/2}(\epsilon +3P)dr\, ,
\end{equation}
while the baryon mass of the static NS is,
\begin{equation}\label{baryonmass}
M_B = \int_0^R 4\pi r^2 e^{\lambda/2}\rho dr\, .
\end{equation}
For the numerical calculation, we use a length scale of $M_\odot
=1$, and the results of our numerical analysis are presented in
Table \ref{table1}. Specifically, in Table \ref{table1}, we
present the maximum baryonic mass, and the maximum gravitational
mass for all the EoS we mentioned earlier, for various values of
the parameter $\alpha$ which is the coupling of the $R^2$ term in
the gravitational action. With regard to the values of the free
parameter $\alpha$, it is expressed in units
$r_g^2=4G^2M^2_\odot/c^4$, see below Eq. (\ref{frdef}). Making the
correspondence with the standard cosmological $R^2$ model, the
small values of $\alpha$ are more compatible with the cosmological
scenarios. However, at this point one must be cautious since the
cosmological $R^2$ model constraints are usually imposed on the
Einstein frame theory. Specifically the constraints on the
parameter $\alpha$ come from the amplitude of the scalar curvature
primordial perturbations, thus yielding a small $\alpha$. If
however the theory is considered in the Jordan frame directly, the
expression of the amplitude of the scalar perturbations is
different compared to the Einstein frame expression, resulting to
different constraints on the parameter $\alpha$. This is why we
chose the parameter $\alpha$ to vary in the range $0<\alpha<10$,
in order to investigate the physics of NSs for a wide range of
values in order to cover the constraints from both frames.
\begin{table}[h!]
  \begin{center}
    \caption{\emph{\textbf{PARAMETERS FOR VARIOUS EOSs}}}
    \label{table1}
    \begin{tabular}{|r|r|r|r||r|r|r|r|}
     \hline
      \textbf{EoS}   & $\alpha$ & $M_B^{max}$ & $M^{max}$ & \textbf{EoS}   & $\alpha$ & $M_B^{max}$ & $M^{max}$
      \\  \hline
        & 0 & 2.65 & 2.17 & & 0 & 2.44 & 2.04\\
      APR & 0.25 & 2.67 & 2.18 &  QHC18 & 0.25 & 2.48 & 2.07 \\
       & 2.5 & 2.76 & 2.24 & & 2.5 & 2.61 & 2.15\\
       & 10 & 2.85 & 2.30& & 10 & 2.70 & 2.22\\
       \hline
        & 0 & 2.47 & 2.08 & &0 & 2.44 & 2.05 \\
      BHF & 0.25 & 2.49 & 2.09 & SLy & 0.25 & 2.46 & 2.06  \\
       & 2.5 & 2.58 & 2.15 & & 2.5 & 2.54 & 2.11 \\
       & 10 & 2.65 & 2.21 & &10 & 2.63 & 2.17 \\
       \hline
        & 0 & 2.84 & 2.38 & & 0 & 2.98 & 2.52 \\
      GM1 & 0.25 & 2.87 & 2.40 & SLy +  & 0.25 & 3.01 & 2.54 \\
       & 2.5 & 3.01 & 2.49 & causal & 2.5 & 3.13 & 2.63 \\
       & 10 & 3.11 & 2.56 & $v_s^2=c^2/3$ & 10 & 3.24 & 2.74 \\
       \hline
       \hline
          \end{tabular}
          \caption{Maximum baryonic mass for stars in $R^2$ gravity for some equation of states and various values of the parameter $\alpha$.}
  \end{center}
\end{table}
Let us discuss the results presented in Table \ref{table1} in some
detail. As a general comment for all the EoSs, the baryonic mass
is significantly larger than the maximum gravitational mass, for
all the values of the parameter $\alpha$, and this is a general
expected result. With regard to the APR EoS, the baryonic mass takes
values in the range $2.65$$M_{\odot}$-$2.85$$M_{\odot}$, and the
maximum gravitational mass in the range
$2.17$$M_{\odot}$-$2.30$$M_{\odot}$. Thus, for the APR EoS, it is
apparent that NSs with baryonic masses larger than
$2.85$$M_{\odot}$ at most, will collapse into black holes. The value
$2.85$$M_{\odot}$ is indicative of the maximum limit of the baryon
mass in the context of $R^2$ gravity and the corresponding
gravitational mass is $2.30$$M_{\odot}$, which means that
astrophysical black holes will be larger than $2.30$$M_{\odot}$ in
the case of $R^2$ gravity and for the APR EoS. With regard to the
BHF EoS, the baryonic mass takes values in the range
$2.47$$M_{\odot}$-$2.65$$M_{\odot}$, and the maximum gravitational
mass in the range $2.08$$M_{\odot}$-$2.21$$M_{\odot}$. Thus, for
the BHF EoS, it is apparent that NSs with baryonic masses larger
than $2.65$$M_{\odot}$ at most, will collapse to black holes. The
corresponding gravitational mass is $2.21$$M_{\odot}$, which means
that astrophysical black holes will be larger than
$2.30$$M_{\odot}$ in the case of $R^2$ gravity and for the BHF
EoS. With regard to the GM1 EoS, the baryonic mass takes values in
the range $2.84$$M_{\odot}$-$3.11$$M_{\odot}$, and the maximum
gravitational mass in the range
$2.38$$M_{\odot}$-$2.56$$M_{\odot}$. Thus, for the GM1 EoS, it is
apparent that NSs with baryonic masses larger than
$3.11$$M_{\odot}$ at most, will collapse to black holes. The
corresponding gravitational mass is $2.56$$M_{\odot}$, which means
that astrophysical black holes will be larger than
$2.56$$M_{\odot}$ in the case of $R^2$ gravity and for the GM1
EoS. With regard to the QHC18 EoS, the baryonic mass takes values
in the range $2.44$$M_{\odot}$-$2.70$$M_{\odot}$, and the maximum
gravitational mass in the range
$2.04$$M_{\odot}$-$2.22$$M_{\odot}$. Thus, for the QHC18 EoS, it is
apparent that NSs with baryonic masses larger than
$2.7$$M_{\odot}$ at most, will collapse to black holes. The
corresponding gravitational mass is $2.22$$M_{\odot}$, which means
that astrophysical black holes will be larger than
$2.22$$M_{\odot}$ in the case of $R^2$ gravity and for the QHC18
EoS. With regard to the SLy EoS, the baryonic mass takes values in
the range $2.44$$M_{\odot}$-$2.63$$M_{\odot}$, and the maximum
gravitational mass in the range
$2.05$$M_{\odot}$-$2.17$$M_{\odot}$. Thus, for the SLy EoS, it is
apparent that NSs with baryonic masses larger than
$2.63$$M_{\odot}$ at most, will collapse to black holes. The
corresponding gravitational mass is $2.17$$M_{\odot}$, which means
that astrophysical black holes will be larger than
$2.17$$M_{\odot}$ in the case of $R^2$ gravity and for the SLy
EoS. Finally, for the theoretical ideal EoS, namely the causal
EoS, the baryonic mass takes values in the range
$2.98$$M_{\odot}$-$3.24$$M_{\odot}$, and the maximum gravitational
mass in the range $2.52$$M_{\odot}$-$2.74$$M_{\odot}$. Thus, for
the causal EoS, it is apparent that NSs with baryonic masses
larger than $3.24$$M_{\odot}$ at most, will collapse to black
holes. The corresponding gravitational mass is $2.74$$M_{\odot}$,
which means that astrophysical black holes will be larger than
$2.74$$M_{\odot}$ in the case of $R^2$ gravity and for the causal
EoS.

Our results hold true for the $R^2$ gravity and for the specific
EoSs which we studied, thus it is conceivable that these are model
dependent. However, it seems that there is a tendency from these
data, that NSs masses cannot be larger than 3 solar masses, and
this combined with the result of Ref. \cite{Astashenok:2021peo},
further supports the GR claim about finding NSs with masses not larger
than 3 solar masses. Also, astrophysical black holes can be found
or created when NSs with baryon masses larger than the
corresponding maximum baryon masses collapse to black holes. In
general, astrophysical black holes can be found even in the lower
limit of the mass gap region, specifically at
$2.5$$M_{\odot}$-$3$$M_{\odot}$, however, our data are model
dependent. Thus our claims are somewhat model dependent and it is
vital to find a universal relation between the maximum baryon
masses and the corresponding maximum gravitational masses in order
to be more accurate. The universal baryon-gravitational masses
relation for $R^2$ gravity can be obtained using the techniques in
Ref. \cite{Gao:2019vby}. The work is in progress along this
research line. This future study will yield a more robust
result and may further indicate, in a refined way, where to find the
maximum masses of NSs and where the corresponding lower masses of
astrophysical black holes.

\section{Concluding Remarks}

In this work we focused on the calculation of the maximum baryonic
mass for static NSs in the context of extended gravity.
We specified our analysis for one of the most important extended
gravity candidate theory, namely $f(R)$ gravity, and we chose one
of the most important models of $f(R)$ gravity, namely the $R^2$
model. We derived the TOV equations for $f(R)$ gravity and
numerically integrated these for the following EoSs, the APR4, the
BHF, the GM1, the QHC18, the SLy, and finally, the limiting case
ideal EoS, called the causal limit EoS. The calculations of the
baryonic mass yielded quite interesting results, with a general
characteristic being that the maximum baryonic mass was higher
than the maximum gravitational mass for the same set of the model's
parameters and the same EoS. The latter characteristic was
expected. However the results tend to indicate some interesting
features for static neutron stars in the context of $R^2$ gravity.
Specifically, the upper limit of all maximum baryon masses for all
the EoSs we studied, and for all the values of the model free
parameters, seems to be of the order of $\sim 3$$M_{\odot}$. This
feature clearly shows that the static NSs maximum
gravitational mass is certainly significantly lower than this
limit, thus the maximum gravitational mass of static NSs
is expected to be found somewhere in the lower limits of the mass
range $2.5$$M_{\odot}$-3$M_{\odot}$. At the same time, one may have
hints on where to find the lower mass limit of astrophysical black
holes, since NSs with baryonic masses larger than the
maximum baryonic mass for the same range of values of the model's
parameters and for the same EoS. Thus one may say that the lower
masses of astrophysical black holes may be found in the same range
$2.5$$M_{\odot}$-3$M_{\odot}$, which is basically the range where
to find the maximum gravitational masses of static NSs.
However, our analysis is model dependent and also strongly depend
on the underlying EoS. Thus, what is needed is to find a universal
 and EoS-independent relation for the baryonic and gravitational
masses of static NSs, at least for the $R^2$ model at hand. The
motivation for this is strong, since we may reach more rigid
answers on the questions which are the maximum NSs masses and
which are the minimum astrophysical black holes masses. This issue
will be thoroughly addressed in the near future. As a final
comment, let us note that it is remarkable that the magic number
of 3 solar masses seems to appear in the context of maximum
baryonic masses. Basically for the baryonic masses, the
$3$$M_{\odot}$ limit is a mass limit that NSs will never reach for
sure. Hence this is an ideal number, not a true limit. On the same
line of research, the causal EoS maximum gravitational mass for
NSs studied in \cite{Astashenok:2021peo}, also involves the
``magic'' number of $3$$M_{\odot}$. Thus the GR claim that neutron
stars cannot have gravitational masses larger than $3$$M_{\odot}$
is also verified from the perspective of baryonic masses
calculations. This conclusion is expected to hold true even if the
universal and EoS-independent relation between that baryonic and
gravitational masses for static NSs is found. However, in order to
be accurate, one must perform similar calculations for spherical
symmetric spacetimes in other modified gravities, like $f(T)$
gravity \cite{Ruggiero:2016iaq,Ren:2021uqb} or even
Einstein-Gauss-Bonnet gravity \cite{Charmousis:2021npl}.

Finally, let us discuss an interesting question, specifically,
whether the analysis we performed would help to break the
degeneracies between the NS EoS and modified gravity forms. Indeed
this would be the ideal scenario and it is generally not easy to
answer. One general answer can be obtained if the following
occurs: if a future observation yields a large mass of a static or
nearly static NSs, which cannot be explained by a stiff EoS, since
the latter is constrained by the GW170817 event. This is seems to
be the case in the GW190814 event, but one has to be certain about
the NSs observation. Thus a kilonova future event, with the
characteristics we described, may indeed verify whether modified
gravity controls eventually large mass NSs and their upper limits.
Our analysis however showed that it is highly unlikely to find NSs
beyond approximately 3 solar masses, and this is an upper bound in
NSs masses. However, we obtained this result in a model and EoS
dependent way, so we need to extend our analysis in a more
universal way. Work is in progress toward this research line.

\section*{Acknowledgments}

This work was supported by MINECO (Spain), project
PID2019-104397GB-I00. This work was supported by Ministry of
Education and Science (Russia), project 075-02-2021-1748 (AVA).
SC acknowledges the support of INFN, {\it Sezione di Napoli, iniziative specifiche} MOONLIGHT2 and QGSKY.

\end{document}